\documentclass[%
 reprint,
 amsmath,amssymb,
 aps,
]{revtex4-1}

\usepackage{graphicx}
\usepackage{dcolumn}
\usepackage{bm}
\usepackage{color}
\usepackage{cases}
\usepackage{here}

\begin{document}
\preprint{APS/123-QED}

%\title{Neural network potential study of point defect properties in multiple charge states: GaN with nitrogen vacancy}
\title{Using neural network potential to study point defect properties in multiple charge states of GaN with nitrogen vacancy}

\author{Koji Shimizu$^1$}
 \email{shimizu@cello.t.u-tokyo.ac.jp}
\author{Ying Dou$^2$}
\author{Elvis F. Arguelles$^1$}
\author{Takumi Moriya$^1$}
\author{Emi Minamitani$^3$}
%\author{Atsushi Kobayashi$^2$}
%\author{Hiroshi Fujioka$^2$}
\author{Satoshi Watanabe$^1$}
  \email{watanabe@cello.t.u-tokyo.ac.jp}

\affiliation{
$^1$Department of Materials Engineering, The University of Tokyo, 7-3-1 Hongo, Bunkyo-ku, Tokyo 113-8656, Japan \\
$^2$Institute of Industrial Science (IIS), The University of Tokyo, 4-6-1 Komaba, Meguro-ku, Tokyo 153-8505, Japan \\
$^3$Institute for Molecular Science, 38 Nishigo-Naka, Myodaiji, Okazaki, 444-8585, Japan 
}

\date{ \today }

\begin{abstract}
\vspace{10mm}
Investigation of charged defects is necessary to understand the properties of semiconductors.
While density functional theory calculations can accurately describe the relevant physical quantities, these calculations increase the computational loads substantially, which often limits the application of this method to large-scale systems.
In this study, we propose a new scheme of neural network potential (NNP) to analyze the point defect behavior in multiple charge states.
The proposed scheme necessitates only minimal modifications to the conventional scheme.
We demonstrated the prediction performance of the proposed NNP using wurzite-GaN with a nitrogen vacancy with charge states of 0, 1+, 2+, and 3+.
The proposed scheme accurately trained the total energies and atomic forces for all the charge states.
Furthermore, it fairly reproduced the phonon band structures and thermodynamics properties of the defective structures.
Based on the results of this study, we expect that the proposed scheme can enable us to study more complicated defective systems and lead to breakthroughs in novel semiconductor applications.
\vspace{10mm}

\end{abstract}

\maketitle

The presence of defects in materials, inevitable in most systems, alters their electronic and dynamic properties from their pristine forms \cite{Oba-APE-2018}.
In semiconductors, as-grown samples often contain certain amounts of native defects, resulting in high carrier concentrations \cite{Monemar-JAP-1979, Bogusl-PRB-1995}.
Moreover, impurity atoms are often introduced intentionally as dopants into materials to control their conductivity.
These dopants can be either $n$- or $p$-type \cite{Ploog-JVSTA-1998}. 
Another problem, particularly in nitride semiconductors, is the inevitability of dislocations during the growth processes of materials.
These dislocations critically impact the mechanical and thermal properties of materials \cite{Mion-APL-2006, Fujikane-PSSC-2010}.
The above discussion highlights the need for research focused on studying defects in semiconductor materials.

First-principles calculations based on the density functional theory (DFT) have been indispensable in studying various defect properties of materials.
The accuracy of such calculations depends on several factors such as the level of approximation in treating the electron-electron interaction, supercell size in modeling, and choice of the functional.
In semiconductors, DFT calculations employing hybrid functionals are known to improve accuracy \cite{Moses-JCP-2011, Gillen-JPCM-2013}; however, this approach increases the computational cost substantially.
Furthermore, performing dynamical calculations, e.g., phonons and molecular dynamics on defective systems, typically requires significant configurational space and long computational times.
These problems highlight the limitations of DFT calculations in treating defect properties even at the local density approximation or generalized gradient approximation (GGA) level.

Interatomic potentials using machine learning (ML) techniques, such as the high-dimensional neural network potential (NNP) \cite{Behler-PRL-2007}, Gaussian approximation potential \cite{Bartok-PRL-2010}, moment tensor potential \cite{Shapeev-MMS-2016}, and spectral neighbor analysis potential \cite{Thompson-JCP-2015}, have been gaining increasing attention because of their low computational costs (by several orders of magnitude) and accuracy comparable with that of DFT calculations.
Such ML potentials have been applied to various materials, e.g., Li$_3$PO$_4$ \cite{Li-JPC-2017}, GaN \cite{Minamitani-APE-2019, Watanabe-JPE-2021}, Au-Li \cite{Shimizu-PRB-2021, Watanabe-JPE-2021}, to successfully obtain the relevant dynamical quantities.

For NN-based ML potentials, several extensions have been proposed to achieve an improved expression of physical quantities.
Interatomic potentials that use a cutoff radius to truncate the surrounding atomic environment in their descriptors generally possess a potential risk of inaccurate descriptions of long-range interactions, particularly when applied to ionic materials.
The initial approach to address this problem was to overlay another NN on the standard one, which was designed for predicting atomic charges and evaluating long-range electrostatic interaction \cite{Artrith-PRB-2011}.
More recently, an advanced method based on using the charge neutrality condition to predict atomic charges demonstrated accurate predictions of non-local charge transfer, which could not be achieved by conventional models \cite{Ko-NC-2021}.

The ML potentials have never been applied to defective systems to treat multiple charge states because of the difficulty in optimizing the fitting parameters owing to the non-unique output values \cite{Zhang-IEEE-2018} (total energies and atomic forces) for the same or similar inputs (structural descriptors).
In this study, we demonstrated for the first time the construction of an ML potential for such systems.
We used wurzite-GaN including a nitrogen vacancy (V$_{\rm N}^q$) as a prototype case, where V$_{\rm N}^q$ takes various charge states $q$ depending on the Fermi level ($E_{\rm F}$).
As mentioned earlier, the defect study of GaN is of significance for many applications, such as light-emitting diodes \cite{Kim-SR-2017} and high-power devices \cite{Mohammad-PIEEE-1995}.

%%%%%%%
\begin{figure}
\includegraphics[bb=0 0 2395 1307, width=0.48 \textwidth]{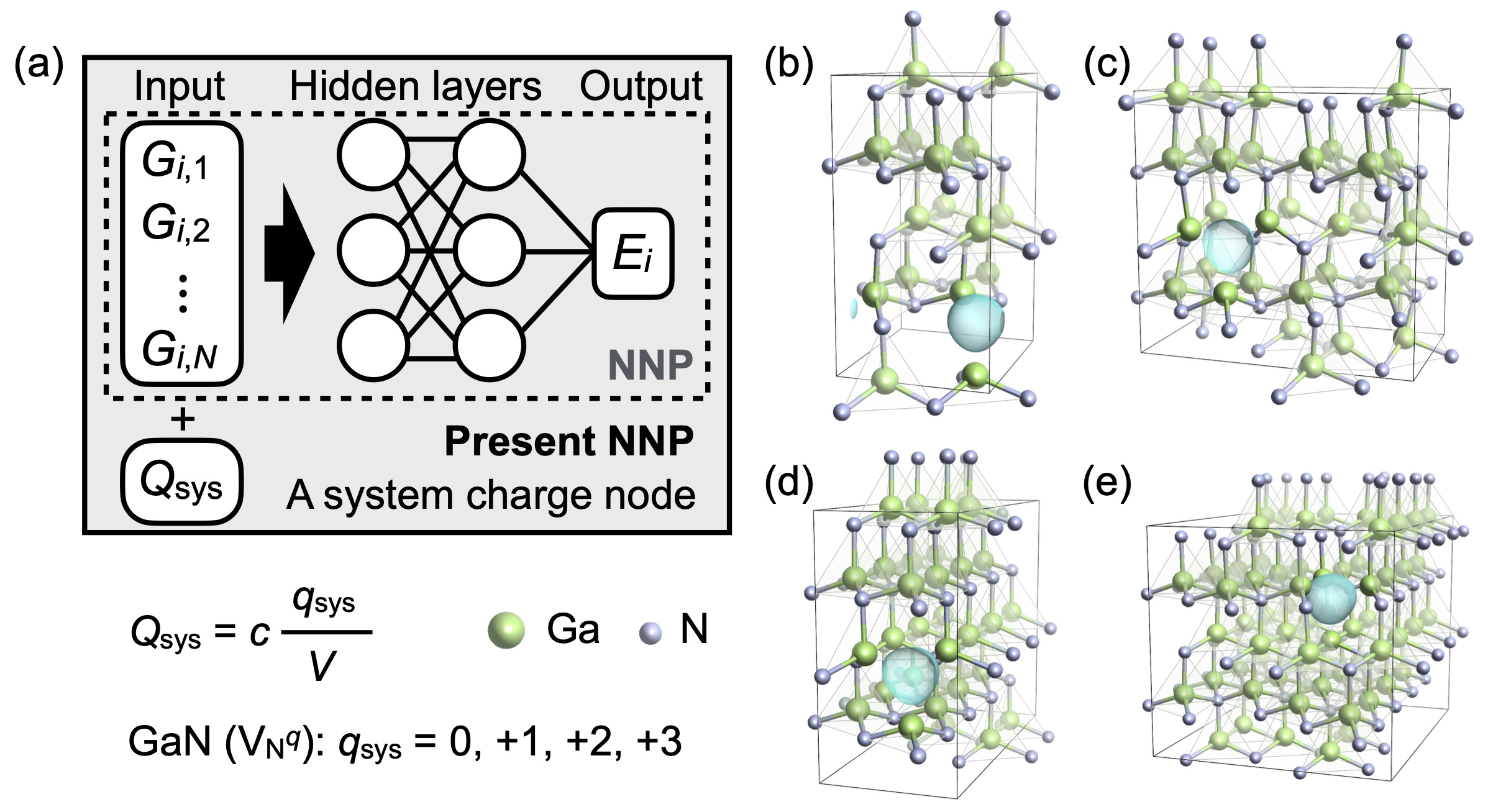}
\caption{
(a) Schematic of the proposed NNP architecture, where the input layer contains the system charge node ($Q_{\rm sys}$).
(b-e) Snapshot of wurzite-GaN structures containing one N-vacancy (V$_{\rm N}$). (b) 31 atoms, (c, d) 63 atoms, and (e) 127 atoms per supercell in total.
The localized defective charge densities are illustrated by the isosurfaces.
Structures are visualized using VESTA package \cite{Momma-jac-2011}.
}
\label{fig1}
\end{figure}
%%%%%%%

To develop our ML potentials, we employed the Behler-Parrinello type NNP \cite{Behler-PRL-2007} as a base model with minimal modifications.
Figure~\ref{fig1}(a) depicts the schematic of the proposed NNP model.
The local atomic features expressed by the radial and angular symmetry functions (SFs) were provided in the input layer.
In addition, we added one system charge ($Q_{\rm sys}$) node in the input layer.
We defined $Q_{\rm sys}$ as constant value $c$ times a charge state of the considered supercell $q_{\rm sys}$ divided by its volume $V$.
For instance, in the cases of pristine and V$_{\rm N}^0$, $Q_{\rm sys}$ = 0.
In contrast, $Q_{\rm sys} = c/V, 2c/V$, and $3c/V$ for V$_{\rm N}^{1+}$, V$_{\rm N}^{2+}$, and V$_{\rm N}^{3+}$, respectively.
Although we used $c = 100$ for the ease of scaling, we emphasize that the present scaling method may not be the optimal one.
All input nodes were fully connected to the first hidden layer.
The atomic energy ($E_{i}$) for the two hidden layer case is expressed as
%%%%%%%%
\begin{equation}
\begin{split}
E_{i} = & f_{a}^{\rm out}[ w_{0, 1}^{\rm out} + \sum_{k=1}^{k_0} w_{k, 1}^{\rm out} f_{a}^{2}\{ w_{0, k}^{2} \\
          & + \sum_{j=1}^{j_0} w_{j, k}^{2} f_{a}^{1} ( w_{0, j}^{1} + \sum_{\mu=1}^{\mu_0} w_{\mu, j}^{1} G_{i}^{\mu} \\
          & + w_{\mu_{0}+1,j}^{1} Q_{\rm sys} ) \} ],
\end{split}
\label{eq1}
\end{equation}
%%%%%%%%
where $G_{i}^{\mu}$ is the SFs of atom $i$ with multiple input nodes $\mu$.
$w_{l, m}^{n}$ denotes a weight parameter connecting the $n$-th and precedent layers of the $l$-th and $m$-th nodes, respectively.
$f_{a}^{n}$ corresponds to the activation function of the $n$-th layer.
We used the hyperbolic tangent for the first and second hidden layers, whereas a linear function was used for the output layer.
The sum of $E_{i}$ over all constituent atoms $E = \sum_i E_i$ is the total energy.
The forces along atomic coordinates $\alpha = x, y, z$ can be expressed as $F_{\alpha_i} = -\partial E / \partial \alpha_i = - \sum_{\mu} \partial E / \partial G^{\mu} \times \partial G^{\mu} / \partial \alpha_i$.
Note that $Q_{\rm sys}$ has no explicit contribution to atomic forces.

To generate the training dataset of NNP, we first performed the molecular dynamics (MD) simulations using the Stillinger-Weber potential \cite{Stillinger-PRB-1985}.
Using the pristine GaN (conventional cell) comprising 32 atoms, we performed canonical ensemble (NVT) MD simulations at temperatures of 300-2700 K with 400-K intervals for 1 ns.
From the obtained trajectories, we extracted the structures of every 10 ps.
We used the optimal lattice constant as well as the compressed and expanded $\pm1$\% and $\pm2$\% and obtained 3500 structures.
Next, we generated V$_{\rm N}$ structures by introducing a nitrogen vacancy into the abovementioned pristine structures.
Using the top two components of the principal component analysis based on the SFs (see \cite{Minamitani-APE-2019} for the parameters of SFs), we performed clustering using a k-means algorithm to generate 3500 clusters.
We chose one data point randomly from each cluster to obtain 3500 V$_{\rm N}$ structures.
In addition, we performed MD simulations with the systems including 64 and 128 atoms in the same manner.
The simulation time was 0.1 ns and the snapshots of every 10 ps were extracted.
In these cases, we obtained 700 and 350 V$_{\rm N}$ structures of 63- and 127-atom systems by randomly introducing a nitrogen vacancy.
We then performed DFT calculations with 0, 1+, 2+, and 3+ charge states for the V$_{\rm N}$ structures.
In this way, we generated 9100 ($3500 \times 2 + 700 \times 2 + 350 \times 2$) structures in total and the corresponding 22750 DFT data.
Figure \ref{fig1}(b-e) depicts one of these V$_{\rm N}$ structures.

We used Vienna Ab initio Simulation Package \cite{VASP1, VASP2} for all DFT calculations.
We used the GGA with the Perdew-Burke-Ernzerhof (PBE) functional \cite{PBE}, plane wave basis set (550 eV cutoff energy), and projector-augmented wave method \cite{Bloechl-PAW}. 
Brillouin zone integration was performed using the sampling technique of Monkhorst and Pack \cite{Monkhorst-Pack} for the training dataset ($4 \times 4 \times 2$ sampling mesh for 32-atom systems, $4 \times 2 \times 2$ and $2 \times 4 \times 2$ for 64-atom systems, and $2 \times 2 \times 2$ for 128-atom systems), whereas the Gamma centered $2 \times 2 \times 2$ was used for phonon calculations.
The convergence criteria were $10^{-7}$ eV and $10^{-4}$ eV/\AA~for self-consistent field and ionic relaxation, respectively.
The number of electrons was varied for V$_{\rm N}^q$ by adding jellium background charges $q$ to neutralize the systems.

%%%%%%%
\begin{figure}
\includegraphics[bb=0 0 2889 1365, width=0.49 \textwidth]{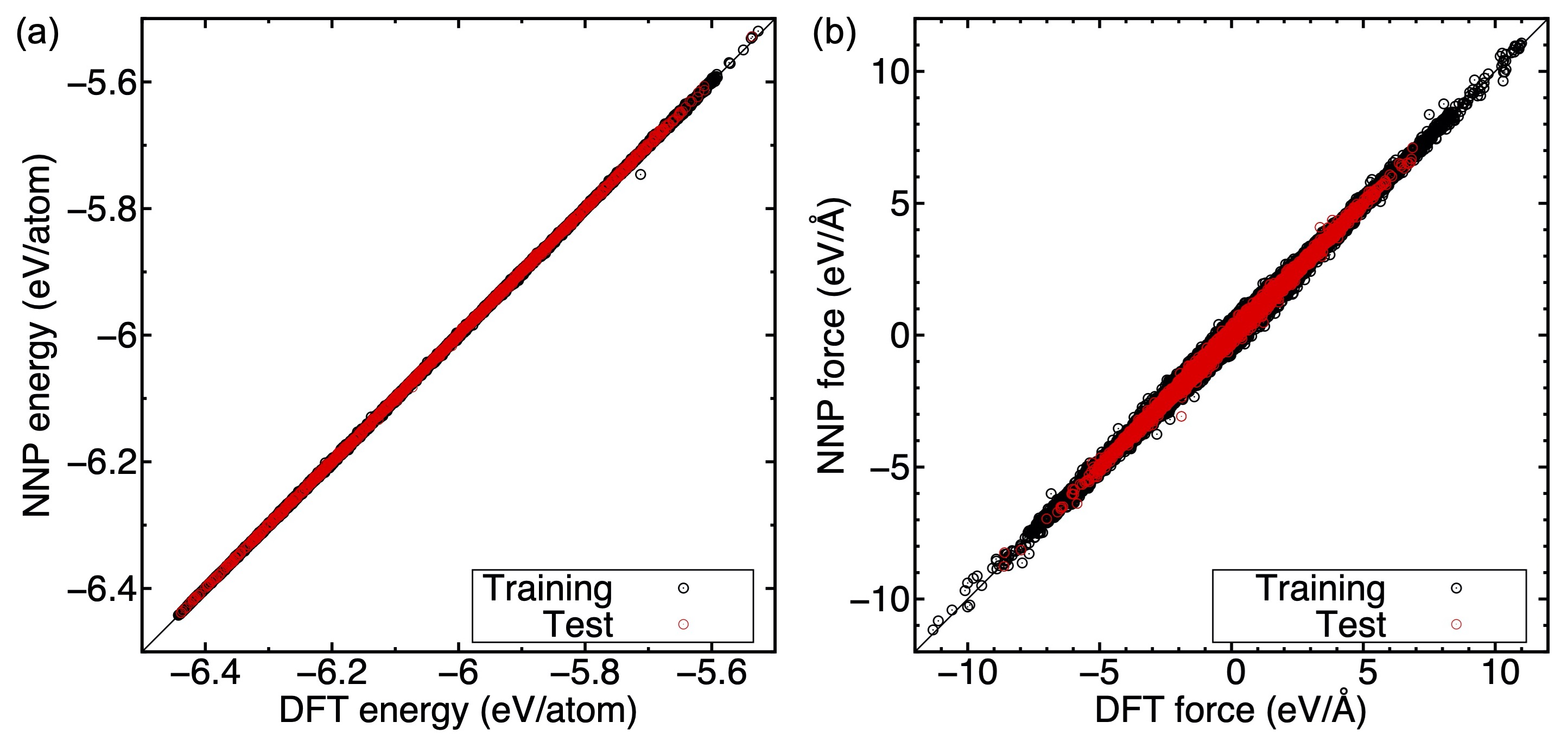}
\caption{
Comparison between DFT and NNP in respect of (a) total energies and (b) atomic forces.
The black and red circles represent the training and test data, respectively.
}
\label{fig2}
\end{figure}
%%%%%%%

%%%%%%%
\begin{figure}
\includegraphics[bb=0 0 3153 3110, width=0.5 \textwidth]{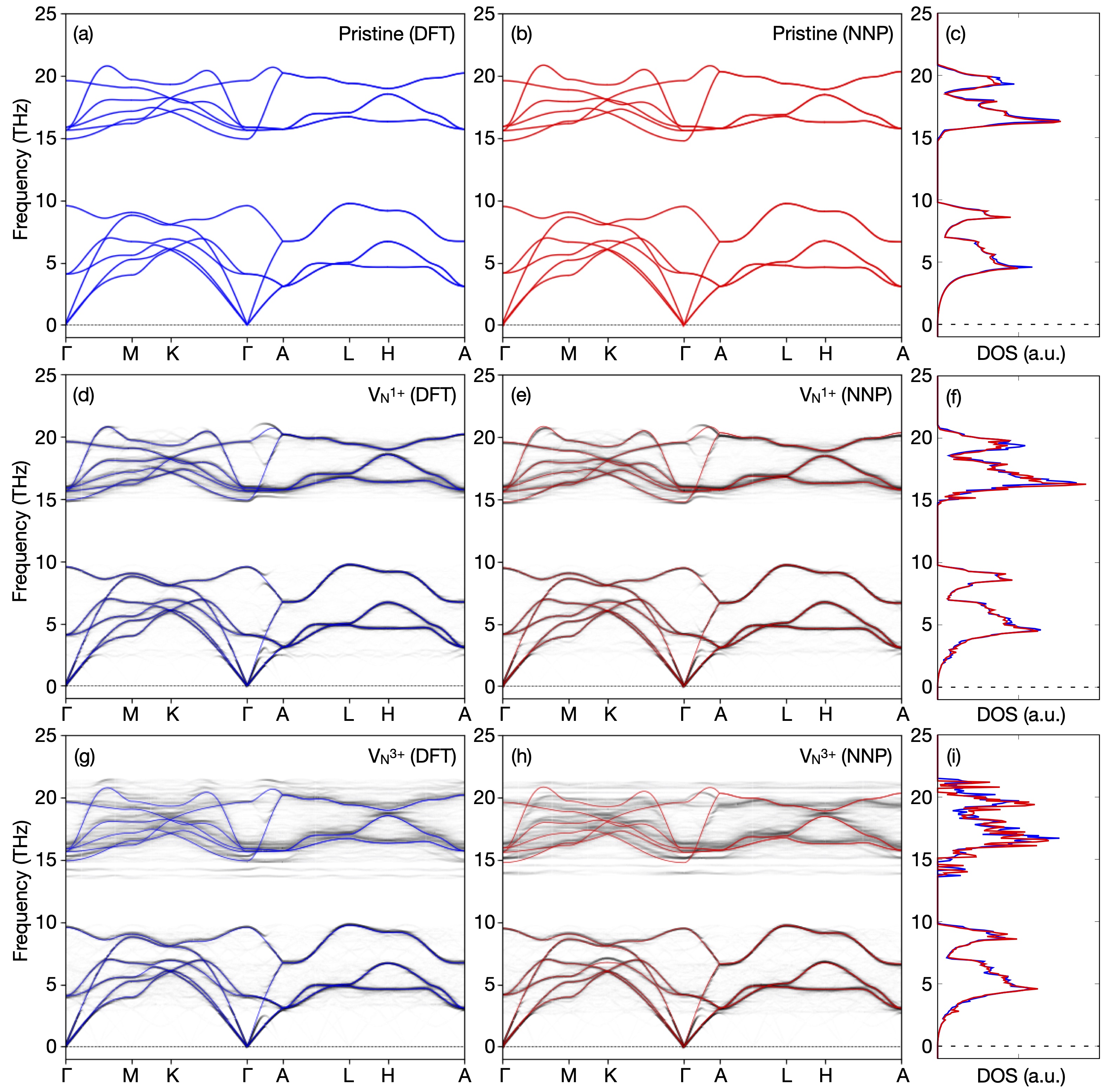}
\caption{
Calculated phonon bands and densities of states of (a-c) pristine, (d-f) V$_{\rm N}^{1+}$, and (g-i) V$_{\rm N}^{3+}$ structures using DFT and NNP.
The densities of states obtained using DFT and NNP are plotted together.
The phonon bands of the pristine case are superimposed on those of V$_{\rm N}^q$.
}
\label{fig3}
\end{figure}
%%%%%%%

Figure~\ref{fig2} depicts a comparison between DFT and the proposed NNP in respect of their total energies and atomic forces.
For this comparison, we used randomly chosen 10\% of the structures from the dataset as the test data.
Both total energies and atomic forces were aligned along diagonal lines, suggesting that the constructed NNP accurately predicted all structures and charge states considered.
The comparison plots using the conventional NNP are given as Fig.~S1 in supplemental material.
The root-mean-square-errors (RMSEs) of the total energies and atomic forces of the proposed model were, respectively, 1.45 meV/atom and 63.8 meV/\AA~for the training dataset and 1.44 meV/atom and 64.6 meV/\AA~for the test dataset.
In this case, we used 8 and 24 types of radial and angular SFs, respectively, for each elemental combination ($8 \times 2 + 24 \times 3 = 88$) with 7 \AA~cutoff distance.
We used the NN consisting of two hidden layers with 30 and 20 nodes.
Therefore, the NN architecture was $[$89-30-20-1$]$.
Note that we used this setting for subsequent calculations based on several trial trainings (see sections 2-4 in supplemental material for the details of the SF parameters, comparison between DFT and the proposed NNP separately depicted for each data type, and transferability tests, respectively).

Figure~\ref{fig3} depicts the calculated phonon bands of the pristine and V$_{\rm N}^q$ structures.
We used a $4 \times 4 \times 2$ supercell with 0.01 \AA~atomic displacement.
The V$_{\rm N}^q$ structures contained one nitrogen vacancy in the surpercell.
All the phonon calculations were performed using $phonopy$ software \cite{Togo-SM-2015}, and the band unfolding package was used \cite{Allen-PRB-2013, XuHe-unfolding}.
In the pristine case, the calculated phonon band structures and densities of states depicted in Fig.~\ref{fig3}(a-c) for DFT and NNP agreed well.
These results were in agreement with experimental results \cite{Ruf-PRL-2001}.
While some slight differences were seen at the higher frequency end, the detailed structures of both acoustic and optical modes matched considerably in the two methods.
Note that we used the optimized structures obtained through DFT for the phonon calculations using NNP.

The calculated phonon band structures and densities of states of V$_{\rm N}^{1+}$ and V$_{\rm N}^{3+}$ are depicted in Fig.~\ref{fig3}(d-f) and (g-i), respectively (see Fig.~S6 in supplemental material for phonon bands of other valence states).
In this case, the pristine phonon bands were superimposed on the defective bands for ease of comparison.
Overall, the introduction of the defect caused band splitting at various regions.
In addition, softening could be seen clearly for the V$_{\rm N}^{1+}$ (also V$_{\rm N}^{0}$) case at around $2 \sim 3$ THz, whereas the band widths of acoustic modes were comparable in all the cases.
We found that the softened mode corresponded to the vibration along the defect position.
In the optimized structure of V$_{\rm N}^0$, the Ga atoms nearest to the defect slightly moved toward the defect direction when compared with their pristine positions. % (0.024 \AA).
In contrast, the Ga atoms moved away from the defect position in the V$_{\rm N}^{1+, 2+, 3+}$ cases, and the shifts were larger for the higher charge states (see Figs.~S7 and S8 in supplemental material for shifts in the Ga atom positions near V$_{\rm N}^{q}$).
A previous study has also reported structural changes in V$_{\rm N}^{1+}$ and V$_{\rm N}^{3+}$ \cite{Diallo-PRA-2016}, which agree with the present results.
This suggests that the potential energy gradient became shallower for the lower charge states, thereby causing the corresponding phonon softening.

The optical phonon modes, on the contrary, differ significantly clearly among the charge states.
Particularly, we found flat bands at the bottom of optical modes in V$_{\rm N}^{3+}$.
The lowest frequency mode corresponded to nitrogen vibrations along the $ab$-plane.
The second lowest mode included nitrogen vibrations along the $c$-direction in addition to the ones along the $ab$-plane.
The phonon band obtained using this NNP also exhibited the emergence of a flat band in V$_{\rm N}^{3+}$.
Furthermore, the proposed NNP reproduced the entire phonon band structures well, although with some slight discrepancies.

%%%%%%%
\begin{figure}
\includegraphics[bb=0 0 2801 1341, width=0.49 \textwidth]{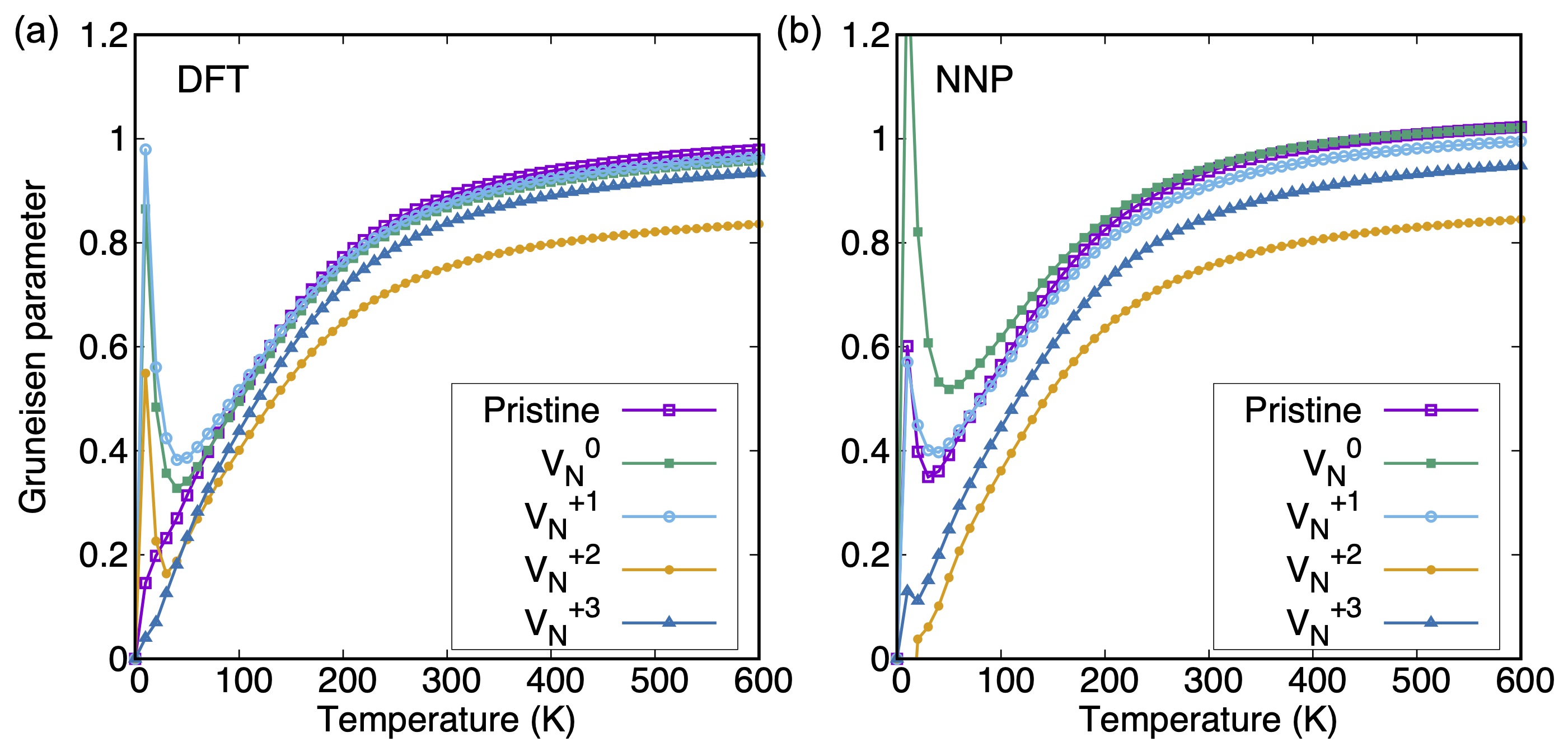}
\caption{
Calculated Gr\"{u}neisen parameters of pristine and V$_{\rm N}^{q}$ structures as functions of temperature: (a) DFT and (b) NNP.
}
\label{fig4}
\end{figure}
%%%%%%%

Further, we evaluated the thermodynamic properties of the defective GaN within the quasi-harmonic approximation.
We used the optimal, $\pm 1$\%, and $\pm 2$\% lattice constants for these calculations.
Figure~\ref{fig4} depicts the calculated Gr\"{u}neisen parameters of the pristine and V$_{\rm N}^q$ systems as functions of temperature.
The obtained values are close to the reported Gr\"{u}neisen parameter of 0.87 at 300 K \cite{Iwanaga-JMS-2000}.
We found that the Gr\"{u}neisen parameters of V$_{\rm N}^{0}$ and V$_{\rm N}^{1+}$ were larger than that of the pristine at $T < 100$ K.
The magnitudes of Gr\"{u}neisen parameters reflect the size of phonon anharmonicity.
Therefore, the V$_{\rm N}^{1+}$ had the strongest anharmonicity among the considered systems.
This result coincided with the observed phonon softening.
In contrast, we found smaller Gr\"{u}neisen parameters for V$_{\rm N}^{2+}$ and V$_{\rm N}^{3+}$, because V$_{\rm N}^{2+}$ had a smaller bulk modulus and thermal expansion coefficient.
For V$_{\rm N}^{3+}$, the thermal expansion coefficient was small; however, its bulk modulus was the largest among all.

The Gr\"{u}neisen parameters obtained using NNP reproduced the DFT results well at $T > 200$ K.
Even though the prediction at $T < 200$ K was not accurate, a qualitative agreement was achieved: the V$_{\rm N}^{0}$ and V$_{\rm N}^{1+}$ (V$_{\rm N}^{+2}$ and V$_{\rm N}^{3+}$) exhibited larger (smaller) Gr\"{u}neisen parameters than that of the pristine case.
Because calculations require a highly accurate prediction performance, we might need DFT calculations or ML potentials with significantly smaller RMSE values as criteria.
Possibly, charged defect corrections for atomic forces may require to be considered.
We leave this aspect for future studies.

%%%%%%%
\begin{figure}
\includegraphics[bb=0 0 2610 1264, width=0.49 \textwidth]{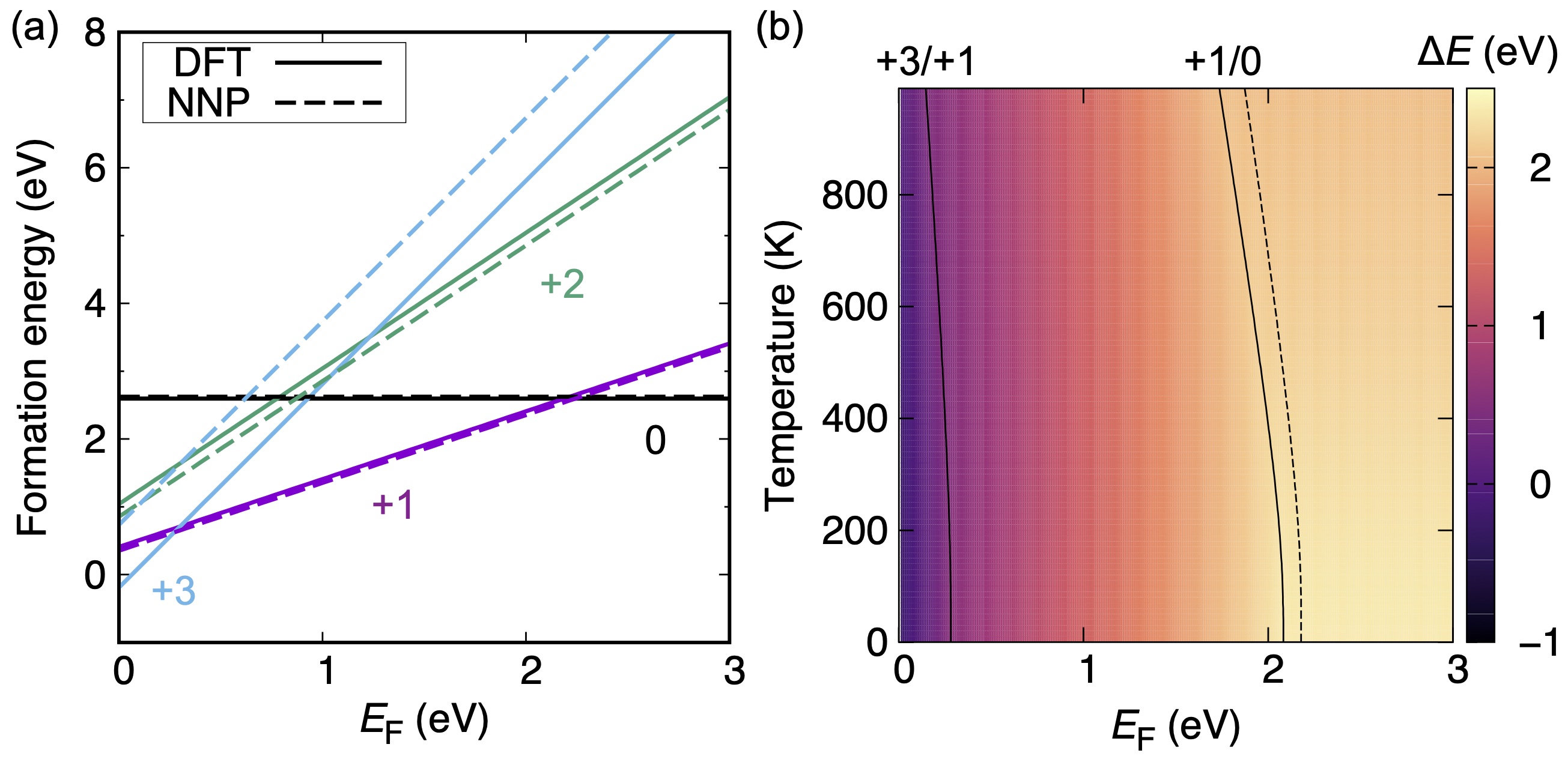}
\caption{
(a) Calculated defect formation energies as functions of $E_{\rm F}$.
(b) Calculated temperature dependence on transition levels of V$_{\rm N}^{q}$.
Solid (dotted) line depicts the DFT (NNP) results.
}
\label{fig5}
\end{figure}
%%%%%%%

Finally, we calculated the vacancy formation energies using the proposed NNP.
In formation energy calculations of charged defects in periodic systems, corrections to eliminate spurious electrostatic interaction energy among the periodic images are necessary.
Various techniques have been proposed in this regard \cite{Freysoldt-PRL-2009, Kumagai-PRB-2014, daSilva-PRL-2021}.
In this study, we did not apply such corrections owing to the inaccessibility of electrostatic potentials by NNP.
However, this problem can be resolved by accounting for corrections as a pre-process for the preparation of training data.

Figure~\ref{fig5}(a) depicts the calculated vacancy formation energies of V$_{\rm N}^q$ as functions of $E_{\rm F}$. 
Note that we considered the Ga-rich condition and used the reference valence band maximum and chemical potential values obtained using DFT for NNP calculations.
Near the valence band side, we obtained the transition from V$_{\rm N}^{3+}$ to V$_{\rm N}^{1+}$ (3+/1+).
In addition, the transition from V$_{\rm N}^{1+}$ to V$_{\rm N}^0$ (1+/0) could be seen at the conduction band side.
These tendencies agreed with those reported in previous studies \cite{Walle-APR-2004}.
The transition levels obtained using NNP reproduced the DFT results well for V$_{\rm N}^{0, 1+, 2+}$.
However, an error of 1.03 eV remained in the V$_{\rm N}^{3+}$ case because the electronic state of V$_{\rm N}^{3+}$ was specifically sensitive to the most stable structure.
The current training dataset generated by adding thermal fluctuations (MD trajectories) could not cover the particular structure-energy relationship.

Figure~\ref{fig5}(b) depicts the temperature-dependent transition levels of V$_{\rm N}^q$ including the phonon contributions.
The color map depicts the Gibbs free energies, and the transition levels are indicated as lines (solid for DFT; dotted for NNP).
In both 3+/1+ and 1+/0, the transition levels decreased when the temperature increased because the phonon contributions (phonon energy and entropy terms) were larger for lower valence states.
The V$_{\rm N}^0$ case had the largest contributions, resulting in the largest shift in the transition level.
The proposed NNP successfully reproduced these shifts, although the 3+/1+ transition was predicted inside the valence band.

In summary, we developed a new scheme of the neural network potential (NNP) to analyze the point defect behavior in multiple charge states.
Our results indicate that minimal modifications to NNP---we added only a system charge node in the input layer of the conventional NNP---can significantly improve the training results of such systems.
We demonstrated the performance of the proposed NNP using wurzite-GaN including a nitrogen vacancy, with 0, 1+, 2+, and 3+ charge states, as a prototype material.
We constructed the proposed NNP using various GaN structures and the corresponding total energies and atomic forces obtained using the density functional theory (DFT) calculations.
The resultant NNP accurately reproduced the phonon band structures and thermodynamic properties of defective systems.
The proposed scheme is expected to pave the way for further advancement in potential applications of ML and growth in materials science research.

Acknowledgements.
This study was supported by
JST CREST Program ``Novel electronic devices based on nanospaces near interfaces" and
JSPS KAKENHI Grant Numbers 19H02544, 19K15397, 20K15013, 20H05285, 21H05552.
Some of the calculations used in this study were performed using the computer facilities at
ISSP Supercomputer Center and Information Technology Center, The University of Tokyo, and Institute for Materials Research, Tohoku University (MASAMUNE-IMR).
We would like to thank Editage (www.editage.com) for English language editing.

\bibliography{apssamp}

\end{document}